\documentstyle[12pt]{article}
\def\lsim{\mathrel{\rlap{\raise 2.5pt \hbox{$<$}}\lower 2.5pt}}
\def\gsim{\mathrel{\rlap{\raise 2.5pt \hbox{$>$}}\lower 2.5pt}}
\begin{document}
\bibliographystyle{plain}
\thispagestyle{empty}
\begin{small}
\begin{flushright}
IISc-CTS-15/97\\
hep-ph/9712525\\
\end{flushright}
\end{small}
\vspace{-3mm}
\begin{center}
{\Large{\bf Chiral Perturbation Theory for Nuclear
Physicists}}\footnote{Invited talk at the DAE 
Nuclear Physics Symposium - 1997,
December 26-30, 1997, Bangalore, India}

\vskip 2cm

{\bf B. Ananthanarayan}, \\
Centre for Theoretical Studies, \\
Indian Institute of Science, Bangalore 560 012, India\\
\[anant@cts.iisc.ernet.in\]\\
\end{center}

\vskip 3cm

\begin{abstract}
Chiral perturbation theory is the low energy effective
theory of the strong interactions for the light pseudoscalar
degrees of freedom.   This program is based on effective
Lagrangian techniques and is an expansion in the powers
of the momenta and the powers of the quark masses, which
correct the soft-pion theorems.  After briefly reviewing
these features and some results, we address the 
implications of this program
to $\pi-N$ scattering, the $\pi-N$ sigma term and some
recent investigations of the implications of
chiral symmetry to 
nucleon-nucleon forces.   We finally look at the implications
of chiral perturbation theory to hadron mass relations.
\end{abstract}

\newpage
\section{Introduction}
The purpose of this talk is to present
in a concise manner some of the results and techniques of
modern chiral perturbation theory\cite{zero,gl1,etc,hl1,jg1}
to highlight the links with nuclear physics.
Much of what I will say here is already
found in standard textbooks\cite{sw1,dgh}, and I will
list several references where indepth
explanations of the subjects can be found.
The pion first posited to explain
the forces between two nucleons is of 
interest to nuclear physics.   The pions $\pi^\pm$
and $\pi^0$ are the lightest hadrons and are
essentially degenerate in mass and lie in an iso-spin
triplet.  Their lightness may be understood 
by regarding them as the Goldstone bosons of 
spontaneously broken axial vector symmetry of
massless QCD; the presence of non-vanishing quark
masses shifting the pion pole to $M_\pi^2=2 \hat{m} B$,
where $\hat{m}$ is the average mass of the u- and d- quarks,
and $B$ is a measure of the vacuum expectation value
$< 0 | \bar{u} u | 0 > =< 0 | \bar{d} d | 0 > =-F^2_\pi B$
and $F_\pi$ is the pion decay constant\cite{gl1}.
QCD is the theory of quark and gluon degrees of
freedom and exhibits the property of asymptotic
freedom in that at large momenta the coupling
constant becomes smaller and one may study the
theory in a perturbation series in the strong
coupling constant $\alpha_S=g^2/(4\pi)$.  The
lagrangian is:  
\begin{equation}
{\cal L}_{QCD}=-{1\over 4 g^2} G^a_{\mu\nu} G^{a\, \mu\nu}+\overline{q}
i\gamma^\mu D_\mu q-\overline{q}{\cal M}q
\end{equation}
If quark mass matrix ${\cal M}$ is set to zero
then the left- and right- chiral projections may
be rotated independently and the quark flavors
rotated amongst each other.  Thus associated with
say $N$ light quark flavors, we have the chiral symmetry
given by the group $SU(N)_L\times SU(N)_R$ (we disregard
the $U(1)_V$ and the anomalous $U(1)_A$ groups.
This chiral symmetry is broken spontaneously to
its vector subgroup $SU(N)_V$, where $V=(L+R)$
and corresponding to the $SU(N)_A$ broken symmetry
we have $N^2-1$ (pseudoscalar) Goldstone bosons.

On the other hand
at low energies, the coupling constant is large
and the problem is intractable.  In particular,
we do not yet know how to obtain the hadronic
spectrum from the QCD lagrangian, nor do we
know the mechanism by which chiral symmetry is
broken spontaneously.   Nevertheless, the QCD
lagrangian provides the justification for the
successful results obtained from PCAC and current
algebra.
The effective low energy theory of the
strong interactions at next to leading order
requires the knowledge of the underlying theory and
the analysis rests of writing down
the generating functional for the currents
of the theory which is the vacuum to vacuum
transition amplitude in the presence of external
sources.  The low energy expansion is one that
involves derivatives of the external sources
a well known example being the Euler-Heisenberg
method of analyzing QED.   
Chiral pertubation theory then is the low-energy
effective theory of the strong interactions and
involves a simultaneous expansion in the mass of
the quarks and the momenta, about the chirally
symmetric $SU(2)\times SU(2)$ limit of the massless
QCD with the the spontaneous breakdown of this
symmetry by the ground state to $SU(2)_V$, the
pions corresponding to the Goldstone bosons of
the broken $SU(2)_A$ generators.
The Goldstone theorem yields
\begin{equation}
< 0|A_\mu|\pi > =F_\pi p_\mu,
\end{equation}
and $F_\pi\approx 93$ MeV.
To leading order, $O(p^2)$, the
effective lagrangian is that of the non-linear
sigma model.  The effective action is
\begin{equation}
Z_1=F^2\int dx {1\over 2} \nabla_\mu U^T \nabla^\mu U
\end{equation}
where $U$ is a four component real $O(4)$
(note that $O(4)\equiv SU(2)\times SU(2)$) unit vector.
This model is not renormalizable
and the loops of the model
lead to divergences which cannot be absorbed into
the parameters of the model.  In order to absorb the
divergences, one is led to introducing higher derivative
interactions, which then allows one to extend the predictions
at leading order in the momentum or derivative to the next order.
The price is the proliferation of coupling constants
that must be extracted from experiment or alternatively
from theoretical considerations such as the behaviour
of the coupling constants in the chiral limit as well
as non-perturbative approaches such as large $N_c$.
The effective lagrangian at $O(p^4)$\cite{gl1} and at $O(p^6)$\cite{fs}
have been worked out.  [When one considers the interactions
of pions with nucleons, the chiral power counting is
different since the pion is now coupled to an external
nucleon.]
For completeness we write down the effective lagrangian at
$O(p^4)$\cite{gl1}:
\begin{eqnarray}
& \displaystyle {\cal L}_4=l_1(\nabla^\mu U^T \nabla_\mu U)^2 + l_2 
(\nabla^\mu U^T \nabla^\nu U)
(\nabla_\mu U^T \nabla_\nu U)+
l_3(\chi^T U)^2+l_4(\nabla^\mu \chi^T \nabla_\mu U)+ & \nonumber \\
& \displaystyle l_5(U^T F^{\mu\nu} F_{\mu\nu} U) +
l_6 (\nabla^\mu U^T F_{\mu\nu} \nabla^\nu U)+ l_7 (\tilde{\chi}^T U)^2+
h_1 \chi^T \chi + h_2 {\rm tr} F_{\mu\nu} F^{\mu\nu} +h_3 \tilde{\chi}^T
\tilde{\chi} &
\end{eqnarray}
where $F_{\mu\nu}$ are covariant tensors involving the external
fields and their derivatives and the vectors $\chi$ and $\tilde{\chi}$
are proportional to the external scalar and pseudoscalar fields.
With this effective lagrangian and with the loops generated by
the non-linear sigma model and appropriate renormalization, one
may obtain the Green's functions of QCD at this order in the momentum
expansion.  At this order, 10 additional coupling constants
enter the effective lagrangian.

Although the number of coupling
constants at $O(p^6)$ are very large ($>100$), those
entering the pion-pion scattering amplitude are still
limited in number.
Of course, once the coupling
constants are fixed from a certain class of experiments,
at that order,  the theory would have predictions
for all other processes at the appropriate level of
accuracy.  Furthermore, the external field technique
permits an off-shell analysis of the Green's functions
of the theory and permits one to study the quark mass
dependence of the Green's functions.

The important processes of $\pi\pi$ and $\pi N$ scattering have
been analyzed in great detail and methods have been described
in standard books\cite{gb,mms,jlp,rgm,bm,gh2}, and will be of
interest to us in this discussion.  Note that
a good deal of the experimental information on the processes
of interest to us has been obtained via dispersion relation
analysis of phase shift information.  In fact, there is
a rich interplay between the effective lagrangian methods
of chiral pertubation theory and dispersion relation theory
which we will describe in some of the following sections.   

In the following sections, we briefly 
review the status of $\pi\pi$ scattering, $\pi-N$
scattering and the ``sigma'' term, 
and implications of chiral symmetry
to hadron mass relations and finally on
the status of nucleon-nucleon forces from
the view point of chiral symmetry.  
A few remarks are listed on other subjects
of interest with some references to the
literature.
Considerably more time is spent on the
first of the topics as it allows us to set up
the framework to discuss the issues raised in
this talk. 
{\it Above all we catalog
a list of references, with some remarks,
from which the reader would get a
glimpse of this vast and fascinating
subject.
}

\section{$\pi\pi$ Scattering}
As an illustration of the techniques used to
obtain several results, we describe the $\pi\pi$
process\cite{mms,gw1} in some detail, although these results
are now nearly 40 years old. 
In axiomatic field theory the validity 
of dispersion relations have
been proved some time ago.  
In the case of $\pi\pi$ scattering dispersion 
relations are particularly simple.
Phase shift information has been analyzed in the past, well 
before chiral perturbation theory or QCD were established.
Today, an analysis that employs chiral results {\it ab initio}
is required to confront experimental data. 
Pion-pion scattering may be described
in terms of a single function $A(s,t,u)$ of
the Mandelstam variables\cite{cm}, $s,t,u$.   
The process is schematically represented by
\begin{equation}
\pi^a(p_1)+\pi^b(p_2)\rightarrow \pi^c(p_3) + \pi^d(p_4)
\end{equation}
and since iso-spin is conserved by the strong
interactions, the transition matrix is given by:
\begin{equation}
A(s,t,u)\delta^{ab}\delta^{cd}+A(t,u,s)\delta^{ac}\delta^{bd}
+A(u,s,t)\delta^{ad}\delta^{bc}
\end{equation}
where the function $A(s,t,u)=A(s,u,t)$ (and is
denoted as $A_s$) due to generalized
Bose statistics and $s=(p_1+p_2)^2$, $t=(p_1+p_3)^2$
and $u=(p_1+p_4)^2$, all momenta taken to be incoming.
If $\sqrt{s}$ represents the centre of mass energy,
then $t$ and $u$ related to the cosine of the centre of mass
scattering angle via $\cos\theta=(t-u)/(s-4)$, $s+t+u=4$
when we set the particle mass to unity.

Since they lie in an iso-spin
triplet, the s-channel amplitudes for definite iso-spin
can be written down:
\begin{eqnarray}
& \displaystyle T^0_s(s,t,u)=3 A_s + A_t + A_u & \nonumber \\
& \displaystyle T^1_s(s,t,u)=        A_t - A_u & \\
& \displaystyle T^2_s(s,t,u)=        A_t + A_u & \nonumber
\end{eqnarray}
which follows from iso-spin coupling.  
This may be rewritten as $M.A$, where 
$$
M=
\left(\matrix{ 3 & 1 & 1 \cr 0 & 1 & -1 \cr 0 & 1 & 1 \cr  } \right)
$$
and $A=[A_s \, A_t \, A_u]^T$.  Note that
$$
M^{-1}=\left(
\matrix{ {1\over 3} & 0 & -{1\over 3} \cr 0 & {1\over 2} & {1\over 2} \cr 0 & 
  -{1\over 2} & {1\over 2} \cr  } \right)
  $$
One convenient representation for dispersion relations
for the amplitudes of definite spin in the t-channel, 
with two subtractions
(convergence is guaranteed with this number of subtractions
as a result of the Froissart bound) is:

\begin{eqnarray}
& \displaystyle
T^I_t(s,t,u)=\mu_I(t) + \nu_I (t) (s-u) + & \nonumber \\
& \displaystyle {1\over \pi} \int_4^\infty {ds' \over s'^2}
\left( {s^2\over s'-s} + (-1)^I {u^2\over s'-u}\right) \sum_{I'} C_{st}^{II'}
A^{I'}_s(s',t)  & 
\end{eqnarray}
where $\mu_I(t), \, \nu_I(t)$ are unknown t-dependent
subtraction constants ($\mu_1=\nu_0=\nu_2=0$), where now $A^I_s(s',t)$
is the absorptive part of
the s-channel amplitude.
The matrix $C_{st}$ is the so-called crossing matrix,
(embodying the fundamental property of crossing in axiomatic
field theory)
the entries of which may be written down from the general
formula resulting from iso-spin coupling in terms of
the Wigner 6-j symbol\cite{fjd} as:
\begin{equation}
C_{st}(c,d)=(-1)^{(c+d)} (2 c + 1)\left\{
\matrix{ 1 & 1 & d \cr 1 & 1 & c \cr} \right\}
\end{equation}

An elementary and alternative derivation of the crossing
matrix is presented below (which the present author has
been unable to trace in the literature) from
considering the simple crossing
relations between $A_s$, $A_t$ and $A_u$ via the matrix
relations:
$$C_{st}=M.B_{st}.M^{-1},$$
where $B_{st}$ is given below which crosses the s- and
t- channels,
$$
B_{st}=\left(
\matrix{ 0 & 1 & 0 \cr 1 & 0 & 0 \cr 0 & 0 & 1 \cr  } 
\right),
$$
which yields for $C_{st}$,
$$
C_{st}=\left(
\matrix{ {1\over 3} & 1 & {5\over 3} \cr {1\over 3} & {1\over 2} & -{5\over 6}
   \cr {1\over 3} & -{1\over 2} & {1\over 6} \cr  } 
   \right).
   $$
Analogously we obtain the matrices $C_{su}$ and $C_{tu}$
(not listed here), by replacing $B_{st}\to
B_{su}$, 
and $B_{st}\to
B_{tu}$ respectively, with      
$$
B_{su}=\left(
\matrix{ 0 & 0 & 1 \cr 0 & 1 & 0 \cr 1 & 0 & 0 \cr  } 
\right) \,\, {\rm and} \, \,
B_{tu}=\left(
\matrix{ 1 & 0 & 0 \cr 0 & 0 & 1 \cr 0 & 1 & 0 \cr  } 
\right) \,  .
$$
Note that we have not listed the relevant expressions involving
6-j symbols for
the s-u and t-u crossing either.

A very convenient form of dispersion relations has been
found\cite{smr1} which eliminates the unknown functions $\mu_I, \, \nu_I$
in favour of the S- wave scattering lengths $a^0_0$ and $a^2_0$,
where the scattering lengths $a^I_l$ arise in the threshold
expansion for the partial wave amplitudes ${\rm Re}f^I_l(\nu)=\nu^l
(a^I_l+b^I_l \nu + ...)$, where the partial wave expansion is
given by $T^I_s(s,t,u)=\sum (2l+1) f^I_l(s) P_l((t-u)/(s-4))$,
$\nu=(s-4)/4$.  This form is:
\begin{eqnarray}
T^I_s(s,t)&=& \sum_{I'}
\mbox{$\frac{1}{4}$}(s\,{\bf 1}^{II'} + t\,
C_{st}^{II'} + u\, C_{su}^{II'})\,T^{I'}_s(4,0)  \\
 &+&\!\!\int_4^\infty \!ds'\,g_2^{II'}(s,t,s')\,A^{I'}_s(s',0)
+\int_4^\infty \!ds'\,g_3^{II'}(s,t,s')\,A^{I'}_s(s',t)\,
.\nonumber
\end{eqnarray}
For our purposes, it is convenient to write the kernels in the form
\begin{eqnarray} 
g_2(s,t,s')&=&-\frac{t}{\pi\, s'\,(s'-4)}\,
(u\, C_{st} + s\, C_{st}\, C_{tu})\left(\frac{{\bf 1}}{s'-t}
+ \frac{C_{su}}{s'-4+t}\right)\nonumber \\
g_3(s,t,s')&=&-\frac{s\,u}{\pi\,s'(s'-4+t)}\left(\frac{{\bf 1}}{s'-s}
+ \frac{C_{su}}{s'-u}\right)\;.
\end{eqnarray}
Furthermore,
$T_s(4,0)=(a_0^0,0,a_0^2).$
 
The property of crossing places constraints on the absorptive
parts of the amplitudes.  However, the presence of 2 subtractions
in these dispersion relations implies that S- and P- waves
do not face any constraints.  It has been shown that the
dispersive representation for the amplitude in the approximation
that only S- and P- waves saturate the absorptive parts of
the amplitudes lends itself to a straightforward comparison
with chiral amplitudes\cite{ab}. 

Nevertheless the higher partial
waves have to be treated with care.  The problem must be accounted
for when we saturate fixed-t dispersion relations with absorptive
parts which are modelled theoretically, perhaps in terms of
resonance propagators, Pomeron exchange, say the Veneziano model, etc.
Alternatively, one may write down dispersion relations in terms
of homogeneous variables\cite{mrw}
which manifestly enforce crossing symmetry;
however there might be a dependence on parameters which parametrize
the curves in the plane of the homogeneous variables on which
the dispersion relations are written down.   While the actual
implications of these constraints may appear academic, present
day chiral computations are seeking to make very accurate predictions
for, say threshold parameters of scattering.  The effective lagrangian
technique produces manifestly crossing symmetric results for the
amplitudes and thus a comparison which is made must ensure that
crossing constraints are respected.  
At $O(p^4)$ 4 additional coupling constants, scale free
coupling constants $\overline{l}_{1,2,3,4}$ enter the
$\pi\pi$ scattering amplitude\cite{gl1}, while we note
that at leading order we have the simple form for the
scattering amplitude
\begin{equation}
A(s,t,u)={s-1 \over 32 \pi F_\pi^2}
\end{equation}
which leads to the prediction for $a^0_0=7/(32\pi F_\pi^2)\simeq
0.16$, (note that $m_\pi=139$ MeV, has been
set to unity).  At $O(p^4)$ the presence of the
infrared singularities of the theory modifies this prediction
substantially and may be expressed in
terms of the four $\overline{l}$'s, in addition to $F_\pi$. 
Estimates for these quantities from disparate sources such
as D- wave scattering lengths [alternatively from $\pi\pi$
phase information directly], $SU(3)$ mass relations and
the ratio of the decay constants $F_K/F_\pi$ gives a
correction of about 25\% to the leading order prediction,
$a^0_0=0.20\pm 0.01$\cite{gl1} (the experimental
value for this number is quoted as $0.26\pm 0.05$\cite{nagels}).  
Note that the scattering
amplitudes in chiral perturbation theory are perturbatively
unitarity; at one-loop order the loops contribute to the
scattering amplitude terms that have the correct analytic
structure corresponding to producing them from the tree
level amplitude by elastic unitarity.

In particular, today the
chiral amplitudes to two-loops have been computed\cite{sternetal,bcegs,gw2}.
The work of \cite{bcegs} in standard chiral perturbation theory
affords an accurate prediction for the parameter $a^0_0$ which
is a soft quantity from the point of view of dispersion relations
and work is in progress to this end.  For some of the latest
information on the experimental as well as theoretical aspects
of the subject, see Ref.\cite{pipiwg}.

The methods of chiral perturbation theory are also of great
use for decays, strong, semi-leptonic, electromagnetic, etc.
One important example where all the methods of effective
lagrangians as well as those of dispersion relations are
of utility are in the decay $\eta\to 3\pi$\cite{etadecay}.

\section{Pion-Nucleon Scattering} 
An excellent introduction to the subjects
of particle and nuclear physics at their
interface is Ref.\cite{rkb}.
The earliest evidence for the existence of
(partially) conserved
currents in the strong interactions came
from the analysis of pion-nucleon scattering.
The Goldberger-Treiman
relation has been since the earliest days,
the cornerstone of
current algebra.  This relation relates the
pion-nucleon coupling constant,
$g_{\pi N}(=\sqrt{4 \pi 13.5}\simeq 13$) to the nucleon axial-vector
coupling constant $g_A(=1.26)$ and $F_\pi$.  
\begin{equation}
g_{\pi N}\simeq {m_N g_A \over F_\pi}
\end{equation}

We have dealt with the example of $\pi\pi$ scattering
at some length, in order to establish the framework for
$\pi N$ scattering, one must account for the differences
in the iso-spin coupling scheme as well as taking into
account the fermion spin.  While the principles remain
analogous, the details vary with regard to crossing principles,
subtractions for the dispersion relations, the partial
wave expansion to name a few aspects.  
The process that is represented by 
\begin{equation}
\pi^a(q_1)+N(p_1)\to \pi^b(q_2) + N(p_2)
\end{equation}
may be represented in terms of four invariant amplitudes
$D^\pm,\, B^\pm$:
\begin{eqnarray}
& \displaystyle T_{ab}=T^+ \delta_{ab}-T^- i \epsilon_{abc} \tau_c
& \nonumber \\
& \displaystyle T^\pm =\overline{u}(p_2) \left[ D^\pm(\nu,t) +{i\over 2m_N}
\sigma^{\mu\nu} q_{2\mu} q_{1\nu} B^\pm(\nu,t) \right] u(p_1) &
\end{eqnarray}
with $s=(p_1+q_1)^2, \, t=(q_1-q_2)^2, \, u=(p_1-q_2)^2, \,
\nu=(s-u)/(4m_N)$.
At a given order, one wishes to compute the 4 invariant
amplitudes to the appropriate orders (which differ by 2).
Note however that since
the pions are the lightest hadrons, the dispersion relations
in that context prove to be the simplest.   The inclusion
of all the aspects of dispersion relations can be easily
found in the literature\cite{gh2}.
The methods of current algebra, PCAC and in the early
days of phenomenological lagrangians, the pion-nucleon
system has been studied in detail.  An excellent introduction
and review of these topics may be found in Chapter 19 of 
Ref.\cite{sw1}.  For instance, we have the lowest order
predictions for the iso-spin 3/2 and 1/2 scattering
lengths of -0.075 and 0.15 which compare favorably with the
data\cite{nagels}.

In the framework
of modern chiral perturbation theory, in a relativistic
framework this process has been first considered in\cite{gss}.
In this work, the authors have extended the analysis of
the Green's functions of QCD with an external nucleon.
Furthermore this environment is an important one for
the tests of chiral predictions\cite{gh,ge}.

More recently the program of heavy baryon chiral
perturbation theory\cite{jm} has been advocated which takes
into account the fact that the nucleon is much more
massive than the pion.  Within this framework there
have been results for the scattering amplitude\cite{ge}. 

\section{$\pi-N$ Sigma term}
The $\pi-N$ sigma term\cite{mes1} is defined as a certain matrix element
that plays a significant role in the testing of chiral predictions
for the pion-nucleon system.  The sigma term is the matrix element:
\begin{equation}
\sigma={\hat{m}\over 2 m_p} <p|\bar{u}u+\bar{d}d|p>
\end{equation}
where $m_p$ is the proton mass and $|p>$ is the physical one-proton 
state.  It value may be inferred from mass relations for the
$\Sigma$-hyperons and the cascades ($\Xi$) and the strangeness
content of the proton.  Furthermore, chiral symmetry implies
that this is also related to one of the pion-nucleon invariant
amplitudes evaluated at an unphysical (but onshell) point,
the Cheng-Dashen point,
which requires the extrapolation of the scattering data.
The comparison of these two completely independent measures
of the sigma term implies a very delicate test of the implications
of chiral symmetry as well as methods of treatment of the
experimental information, which is why it is a subject of
great interest\cite{gh}.
Furthermore, it behaves as a probe
of the contribution of the strange sea to the proton mass and
as a result is a quantity worthy of considerable theoretical
study.

\section{Nucleon-Nucleon Forces}
Recent work on the application of chiral symmetry
to nucleon-nucleon interactions has been spurred by
proposals by Weinberg\cite{sw2,orvk}.  
The approach taken here extends the ordering of
diagrams in field theory by numbers of powers of
soft pion momenta to processes involving nucleons
as well. 
A guide to the description of forces from 
chiral viewpoint is to enforce the Adler consistency
condition on scattering amplitudes which may be used
to rule out of modify models.
In
Ref.\cite{ksw}, an effective field theory approach
to compute specific scattering amplitudes has been
pursued.  This involves a modified expansion which
does not correspond to an expansion in $m_\pi$.
A different approach to the problem has been advocated in
Ref.\cite{ahs};
it is observed that loop effects dominate over
tree level effects and that small explicit breaking of chiral
symmetry controls long range attraction
in the scalar-isoscalar channel for $NN$ forces.
Analogous observations have been made for multinucleon forces
which differ from other conclusions\cite{sw2,orvk}.

\section{CHPT and Hadron Mass Relations}
The presence of non-analytic corrections to the mass spectrum
of the hadrons is an important subject which leads to the
estimates for the quark mass\cite{jg1}.  In the absence of
iso-spin breaking, one has at tree level algebraic relations
between the masses of the pseudoscalar octet mesons
which yields ratios for the quark masses in terms of the
mass ratios of the pseudoscalar masses.  Details of the
spectrum are then fed in to extract the corrections.
Analogously iso-spin breaking is then treated to obtain
the mass splitting between the u- and d- quarks in terms
of the mass splitting of the pion triplet.  This is
then extended to the baryon sector as well.  Well known
examples in the quark model are the Gell-Mann Okubo mass
formula for the octet and the baryonic sector.  More
recently in heavy baryon chiral perturbation theory, these have
been revisited as well\cite{borasoy,mkb}.  

\section{Other Avenues}
The chiral principles being as important as they are
in the study of strong interaction physics can be
used as a guide for physics outside the realm of
perturbation theory as well.  An interesting application
of chiral principles has been recently proposed\cite{jp}
in order to study the behaviour of pion amplitudes
in the nuclear medium.  Combining the chiral principle
of the presence of an Adler zero with the Veneziano model,
implies
a modification of the Regge slope for the the dual resonance
formula of the scattering amplitudes.  This approach
is found to imply overall consistency between experimentally
observed features such as the drop in the $\rho$ mass,
$\Delta-N$ mass difference, etc., in terms of a universal
parameter.  In the heavy fermion formalism of chiral
perturbation theory for meson exchange currents in nuclei
has been developed and applied to nuclear axial charge
transitions\cite{pmr}.
Interesting issues are discussed in Ref.\cite{tk} regarding
the possible discrepancy between the results of chiral
perturbation theory and those of PCAC-current algebra approach
in dense matter.  For a discussion on the ncleon and strongly
interacting material, see Ref.\cite{vs}.
In the context of nucleon-nucleon potentials and
chiral symmetry, see. Ref.\cite{rm}

\section{Epilogue}
Chiral perturbation theory is the effective low energy
theory of the strong interactions which takes into account
the known symmetry structure of the QCD lagrangian and the
properties of the spectrum.  In particular, the interactions
of the mesonic sector are well understood while the baryonic
sector offers theoretical challenges.  We are in an era when nuclear
physics takes all these inputs into account to finally produce
a framework for nucleon-nucleon forces and interactions.

\vskip 2cm 
\noindent {\bf  Acknowledgements:}  It is a pleasure to thank
Drs. P. B\"uttiker, N. D. Hari Dass, J. Pasupathy and
M. Sainio for discussions.  I also thank Drs. B\"uttiker
and Sainio for reading the manuscript and making invaluable
suggestions.                 

\newpage
\begin{small}

\end{small}

\end{document}